\pdfoutput=1
\documentclass{article}
\usepackage{graphicx}
\usepackage{amsmath}
\usepackage{booktabs}
\usepackage{threeparttable}
\usepackage{multirow}
\usepackage[margin=0.6in]{geometry}  
\usepackage{authblk}
\usepackage{cite}
\usepackage{float}

\newcommand{\bftab}{\fontseries{b}\selectfont}
\renewcommand{\arraystretch}{1.5}  
  
\providecommand{\keywords}[1]{\textbf{Key words:} #1}

\title{Asset Pricing and Deep Learning}
\author{Chen Zhang}
\affil{SenseTime Research, Shanghai, China\\ demi6d@gmail.com}

\date{\today}

\begin{document}

\maketitle

\begin{abstract}
Traditional machine learning methods have been widely studied in financial innovation.
My study focuses on the application of deep learning methods on asset pricing.

I investigate various deep learning methods for
asset pricing, especially for risk premia measurement.
All models take the same set of predictive signals (firm characteristics, 
systematic risks and macroeconomics).
I demonstrate high performance of all kinds of state-of-the-art (SOTA) deep learning
methods, and figure out that RNNs with memory  mechanism and attention have the best
performance in terms of predictivity.
Furthermore, I demonstrate large economic gains to investors using deep learning forecasts.

The results of my comparative experiments highlight the importance of domain knowledge and
financial theory when designing deep learning models. 
I also show return prediction tasks bring new challenges to deep learning.
The time varying distribution causes distribution shift problem, which is essential for
financial time series prediction.

I demonstrate that deep learning methods can improve asset risk premium measurement.
Due to the booming deep learning studies, they can constantly promote the study of
underlying financial mechanisms behind asset pricing. 
I also propose a promising research method that learning from data and 
figuring out the underlying economic mechanisms through explainable artificial intelligence
(AI) methods.
My findings not only justify the value of deep learning in blooming fintech development,
but also highlight their prospects and advantages over traditional machine learning methods.
\end{abstract}


\section{Introduction}
In this paper, I mainly build on the work of \cite{gu2020empirical}, and perform a 
comparative analysis of various SOTA deep learning methods for asset pricing. 
I focus on asset risk premia measurement problems, which are essential in asset pricing.

My primary contributions are extending the study of traditional machine learning methods
to the all kinds of SOTA deep learning methods on financial problems.
And with the empirical comparative analysis, I gain several new findings
in the interdisciplinary study of deep learning methods and asset pricing problems.

\subsection{What is Deep Learning}
Formally, deep learning is a subfield of machine learning, so it inherits all the theories
of machine learning.
Due to its high performance, wide range of applications and uniformed architectures
(neural network), deep learning is commonly studied as a separate field from traditional
machine learning methods such as logistic regression, SVM, tree based models, etc. 
For clarity, I will use machine learning referring to traditional machine learning
in this paper.

Connectionism known as connectionist networks or artificial neural networks (NNs) is
an approach to the study of human cognition that utilizes mathematical models.
Deep learning is exactly the combination of machine learning and connectionism.
In recent years, deep learning methods as mainstream in machine learning studies is 
blooming and gain large success in AI literature.

\begin{figure}[H]
\centering
\includegraphics[width=14cm]{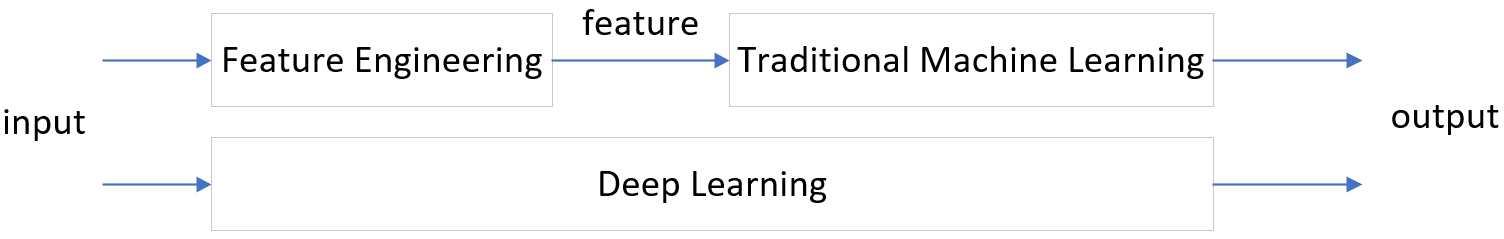}
\caption{Deep Learning}
\textnormal{Deep Learning can automate the feature engineering process.}
\label{fig:deep learning}
\end{figure}

Machine learning is essentially learning a prediction function from input data to output, 
and deep learning is essentially learning various data representation throughout
such process.
Data representation learning is also regarded as automatic feature engineering 
which usually is manual in machine learning.
Figure \ref{fig:deep learning} show the contrast between machine learning and deep learning.

Deep learning is good at finding high dimensional, nonlinear and deep relationships.
Domain specified deep learning models such as Convolutional Neural Networks (CNNs) and
Recurrent Neural Networks (RNNs) are good at processing non-structure data.

\subsection{Why Apply Deep Learning to Asset Pricing}
Empirical asset pricing is essentially a dynamic prediction problem,
and the input (predictive signals) and output (asset risk premium) vary over time.
Since there is a large amount of factors related to stock returns, 
the dimension of inputs (predictors) is usually high.
The relationship between input information and output returns is also probably nonlinear.

Deep learning has its special advantages over machine learning in asset pricing problems.
Machine learning is hard to take advantage of the economic theory and introduce it into
the model as prior knowledge.
However, deep learning can design specific NNs architecture suitable to different scenarios. 
Therefore, it's more flexible and with more spaces for improvement.
For example, sequence models are proper kind of architectures for time series data
and time varying prediction.

Deep learning also called deep neural network, because they are loosely inspired
by biological neuroscience. 
One essential reason deep learning gain great success in the AI industry is that
it takes advantage of the greatest art (brain mechanisms) of nature.

What's more, end to end type deep learning models can avoid tedious work of
feature engineering in machine learning.

\subsection{What Specific Deep Learning Methods Do I Study}
I select a set of typical deep learning methods which gain tremendous success in 
all fields of AI applications. 
They are representative for all kinds of SOTA deep learning models.

This includes deep feedforward NNs such as DNNs, Residual DNNs 
(i.e., DNNs with skip connection), CNNs and residual CNNs. 
This also includes sequential models such as RNNs, Long Short-Term Memory (LSTM),
Gated Recurrent Unit (GRU) and RNNs with attention mechanisms.
At last, I introduce the Transformer, whose NNs architecture is entirely based on
attention mechanism, which has achieved better performance than all kinds of RNNs
on various Natural Language Process (NLP) tasks.

\subsection{Main Empirical Findings}
Refer to the work of \cite{gu2020empirical}, I perform a large scale empirical analysis,
investigating 30,000 individual stocks over 40 years from 1976 to 2016. 
Furthermore, I conduct my analysis with 2 kinds of time span samples,
long time span (40 years) and short time span (5 years).
My predictor set include 81 characteristics for each stock,
6 macroeconomic proxies and 5 factors by \cite{fama2015five} as proxies of systematic risks.

I gain several new findings through my empirical analysis.

All SOTA deep learning methods have high performance in stock return prediction,
and far outperform the traditional linear prediction model.
It demonstrates the great promise of deep learning method for asset pricing.

RNNs with memory mechanism and Transformer have the best performance,
and this means history data has predictive power for asset return.
RNNs are good at merge past and present information together to make prediction.
The memory mechanism attention and cell unit in RNNs both works well, and
long term memory improves stock return prediction.

CNNs have relative worse performance than other NN models, and
transformer is not completely dominating all other RNNs.
It shows the importance of domain knowledge and theory guided model design.
There is no free lunch, and specific model design is essential for problems in different
scenarios.

Skip connections with deep layers have only a little improvement, and middle or shallow
networks still work well.
It shows the fundamental associations between market data and asset prices is not
complex comparing with other deep learning applications,
and studies of economic mechanism behind asset pricing is promising.

Short time span samples have better performance.
This demonstrates the distribution of stock data varies over time, 
and the closer the information, the more predictive power. 
Therefore, distribution shift problem arises in stock returns prediction.

\subsection{What Deep Learning Cannot Do}
Deep learning is good at predicting excess return condition on market information.

However, since the complete theory of deep learning models is still ongoing,
the shortage of deep learning is just like most machine learning methods that
it's hard to fully explain the mechanisms behind the model.

Deep learning models are designed only with reference to,
but not entirely based on, economic mechanisms.
While in this respect, deep learning is still better than machine learning because I can
take advantage of economic mechanisms when designing the NNs' architecture.
I can take the prior knowledge of financial theory as induction bias of the model for
corresponding financial problems.
Also, some studies aim at mitigating this shortage in deep learning literature
such as explainable AI.

\subsection{Literature}
My work extends the literature on empirical asset pricing in several aspects.

First, I extend the machine learning based empirical asset pricing, such as stock pricing
and bond pricing studied by 
\cite{gu2020empirical}, \cite{LEIPPOLD2021} and \cite{bianchi2021bond}.
While deep learning methods talked in this paper are not limited to
specific kind of asset return prediction. 

Second, I extend machine learning based cross-section stock returns study
such as factors dimension reduction by \cite{gu2021autoencoder}. 

Third, I extend the time series return prediction, what is surveyed by 
\cite{koijen2011predictability} and \cite{rapach2013forecasting}.

There are many deep learning methods have appeared in the asset pricing literature.
Some papers are about derivatives pricing via NNs such as
\cite{hutchinson1994nonparametric} and \cite{yao2000option}. \\
\cite{sirignano2016deep} use deep learning to measure mortgage risks, and \\
\cite{heaton2016deep} use deep learning for portfolio management.

My paper focus on studying various SOTA deep learning methods,
and comparing their results on stock return prediction tasks.
I conduct a deep fundamental analysis on the pattern and relationship lie behind 
the market data and stock returns.

\section{Methodology}
In this section, I give a brief description of all the deep learning methods
in my analyzing list.
Details of most basic deep learning mechanisms are well illustrated by
\cite{goodfellow2016deep}.

I use the same objective function for all models which is Mean-Square Error (MSE).
I use learning algorithms substantially based on gradient decent,
which is the most common optimization algorithm for deep learning. 
In particular, I choose Adam, the enhanced gradient decent learning algorithm,
as my basic optimizer, and make some additional improvements such as
batch normalization and layer normalization on it.
I mostly use Dropout as the regularization method in all of my deep learning models.

I use the same asset pricing model as described by \cite{gu2020empirical},
an additive prediction error model for excess return.

\begin{equation}
    r_{i, t+1} = y_{i, t+1} = E_{t}(y_{i, t+1})+\epsilon_{i, t+1},
\end{equation}
where
\begin{equation} \label{eq:over-arching}
    E_{t}(y_{i, t+1})=f(x_{i, t};\theta^{\star})
\end{equation}

\subsection{Objective Function}
For all of my deep learning models, I estimate the parameters through 
Max a Posterior Estimation (MAPE).

I want to maximize the posterior probability of $\theta$ given the samples of
$X$ and $Y$.
\begin{equation}
    \boldsymbol y = f(x; \theta) + \boldsymbol \epsilon
\end{equation}
\begin{equation}
    E(\boldsymbol y) = f(x; \theta)
\end{equation}
my 2 assumptions are that the noise $\epsilon$ follows a normal distribution,
and the prior distribution of parameters $\theta$ is uniform.
\begin{equation} \label{eq:1}
    \begin{aligned}
        \theta^{\star} & = \mathop{\arg\max}_{\theta} P(\theta|\boldsymbol Y, X)\\
        & = \mathop{\arg\max}_{\theta} 
        \frac{P(\theta)P(\boldsymbol Y|\theta, X)}{P(\boldsymbol Y|X)}\\
        & = \mathop{\arg\max}_{\theta} P(\boldsymbol Y|\theta, X)
    \end{aligned}
\end{equation}
Then MAPE is the same as Maximum Likelihood Estimation (MLE) of $\theta$
given samples $(X, Y)$.
Have the Gaussian noise assumption, and take the natural log of equation (\ref{eq:1}), 
MLE is equal to minimizing MSE.
Therefore, my objective function for the model is to minimize the MSE, 
so the loss function $\mathcal{L}(\theta)$ of my deep learning models is defined as MSE.
\begin{equation}
    \begin{aligned}
        \mathcal{L}(\theta) & = \|Y - f(X;\theta)\|^{2} \\
        & = \frac{1}{N \cdot T} \sum_{i=1}^{N} \sum_{t=1}^{T}
        \left(r_{i, t+1}-f\left(x_{i, t} ; \theta\right)\right)^{2}
    \end{aligned}
\end{equation}

\subsection{Learning Algorithm}
Adam introduced by \cite{kingma2014adam} is an algorithm for
first-order gradient-based optimization.
Adam is an algorithm putting Momentum and RMSprop mechanisms together.

Batch Normalization and Layer Normalization are 2 kinds of data normalization method
proposed respectively by \cite{ioffe2015batch} and \cite{ba2016layer}.
The idea of these mechanisms is normalizing the input and hidden unit data, so that
different data are at the same scale.
They can flatten the gradients during the gradient decent process.

Regularization is used to improve the generalization ability of deep learning models.
It makes the model effectively generalize to out-of-sample data set
instead of overfitting the training samples.

In each training batch, overfitting can be significantly reduced by ignoring part of the
feature detectors (leaving part of the hidden layer nodes with values of 0).
This mechanism works in 2 aspects.
First, it reduces the network connections and parameters, thus simplify the neural network.
Second, it implicitly assembles various subnetworks to get the final model of
more generality.

I divide the data set across the time by ($80\%, 10\%, 10\%$), respectively
as train set, validation set and test set.
Train set is used to train the models, and estimate the parameters.
Validation set is used to tune all the hyperparameters of the models.
It can prevent overfitting and ensure the best performance for out-of-sample data.
Test set is only used to test the final out-of-sample performance of the models.

\subsection{Simple Linear}
I take the traditional regression model, linear regression, as my baseline.
\begin{equation}
    f\left(x_{i, t} ; \theta\right)=x_{i, t}^{\prime} \theta
\end{equation}
I also use the most common estimation method, ordinary least squares (OLS),
to estimate the parameters.
\begin{equation}
    \hat{\theta}_{OLS} = (X^{\prime}X)^{-1} X^{\prime}Y
\end{equation}

\subsection{Deep Feedforward Networks}
The idea of Deep Neural Networks (DNNs) is first introduced by
\cite{ivakhnenko1971polynomial}.
The core idea of DNNs is representation learning.
Each unit of one layer can be seen as a combination of previous layer's units,
so it's essentially a modularizing process.

From mathematic perspective, DNNs is a sequence of transformations of vectors.
Each transformation (layer) is composed of affine transformation followed by a nonlinear
function (i.e., activation function).
Deep feedforward networks are often called Multilayer Perceptron (MLP).
MLP with one hidden layer can approximate any function $f(x)$.
MLP is the standard DNNs, and I will call it DNNs for simplicity in this paper.
The special structure of DNNs also contains an implicit L2-norm Regularization effect.
Most DNN models are called feedforward, because the information flow from input to output in
one-way direction without feedback. 
Models with feedback connections are called RNNs.

DNNs are essentially approximating a function from $x$ to $y$:
\begin{equation}
    y = f^{\star}\left(x \right)
\end{equation}
\begin{equation}
    y_{i, t} = f^{\star}\left(x_{i, t} \right)
\end{equation}
\begin{equation}
    y_{i, t} = f\left(x_{i, t} ; \theta^{\star}\right)
\end{equation}
Every layer in DNNs consists of two components, affine transformation and activation 
function.
\begin{equation}
   \boldsymbol z = \boldsymbol W^{\prime}\boldsymbol h + \boldsymbol b
\end{equation}
The activation function used in my model is ReLU proposed by \cite{glorot2011deep}.
\begin{equation}
   ReLU(\boldsymbol z) = max\{\boldsymbol 0, \boldsymbol z\}
\end{equation}
The total data transformation in each DNN layer is as follows:
\begin{equation}
   \boldsymbol h^{<i+1>} = ReLU(\boldsymbol W^{<i+1>\prime}\boldsymbol h^{<i>}
   + \boldsymbol b^{<i+1>})
\end{equation}

\subsubsection{Residual DNNs}
Instead of approximating a function, each layer in Residual DNNs approximates
the residual between input and output of this layer. 
When I introduce the deep residual neural network framework, 
I explicitly let the layers fit a residual function.
\begin{equation}
    f(x) := h(x) - x
\end{equation}

\begin{figure}[H]
\centering
\includegraphics[width=8cm]{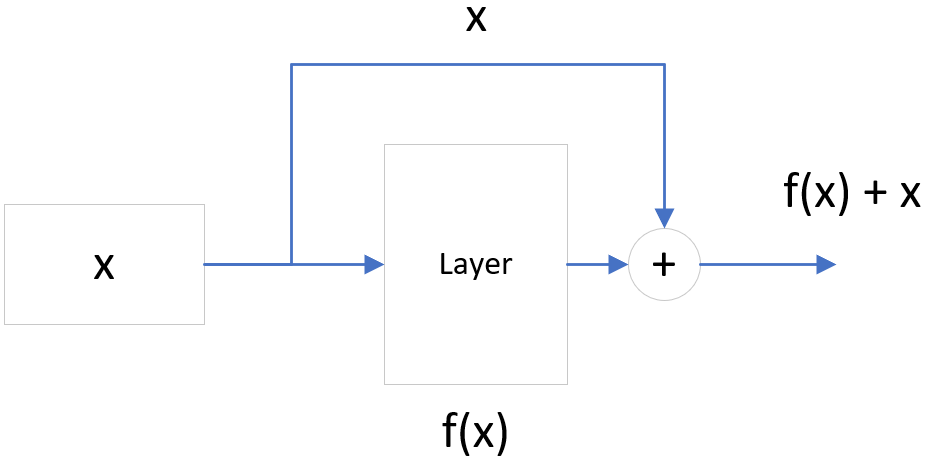}
\caption{Residual DNNs}
\textnormal{Layer of Residual DNNs approximates the residual between input and output.}
\label{fig:Residual DNNs}
\end{figure}

One implementation of residual networks is standard DNNs with a skip connection
between each layer.
Skip connection connects layer $i$ to layer $i + 1$ or higher layer is shown in Figure
\ref{fig:Residual DNNs}.
Skip connections make it easy for the gradient to flow back during the backward propagation.
Skip connections between layers reduce the length of path from lower layer to
the output, and thus mitigate the vanishing gradient problem.

\subsection{Convolutional Networks}
\begin{figure}[H]
\centering
\includegraphics[width=11cm]{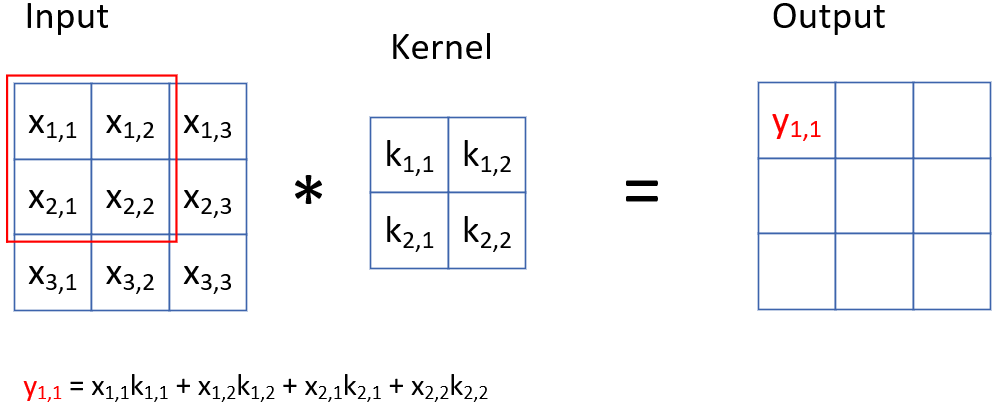}
\caption{Convolution}
\textnormal{The convolution layer performs a convolution operation on inputs with the kernels.}
\label{fig:Convolution}
\end{figure}

CNNs have been demonstrated well performance in image recognition by
\cite{krizhevsky2012imagenet}.
In deep learning literature, I regard the cross-correlation function as convolution,
while formally cross-correlation stands for convolution without flip the kernel. 
\begin{equation}
    S(i, j)=(I * K)(i, j)=\sum_{m} \sum_{n} I(i+m, j+n) K(m, n)
\end{equation}
Figure \ref{fig:Convolution} shows the convolution operation.

The core idea of CNNs is that kernels in convolution layer stand for the receptive field.
Such architecture design has two advantages.
First is sparse connectivity, that neurons only have connections within receptive field.
Second is parameter sharing, each kernel has the same weights moving across all the
inputs of the layer.

\begin{figure}[H]
\centering
\includegraphics[width=11cm]{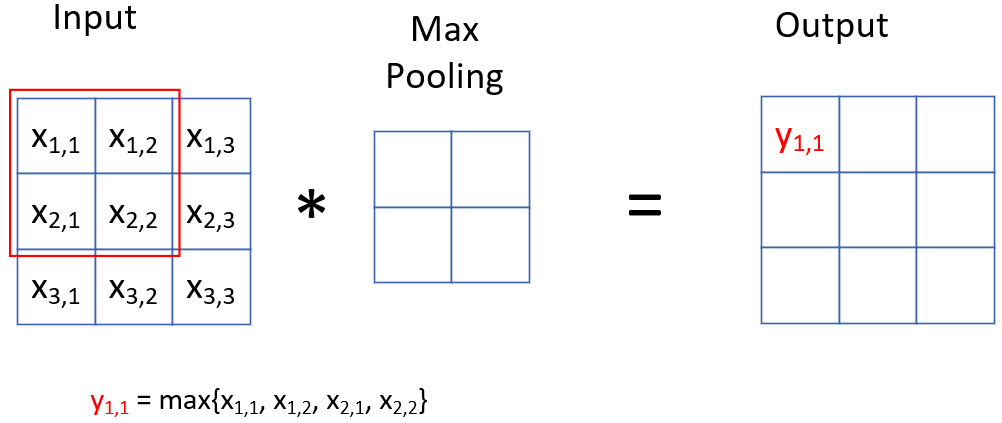}
\caption{Pooling}
\textnormal{Max pooling layer take the maximum value of inputs within the kernel range.}
\label{fig:Pooling}
\end{figure}

Another special layer type in CNNs is Pooling.
It essentially is a summary statistic of each layer.
There are different kinds of pooling layers, such as max pooling and avg pooling.
Max (or avg) pooling takes the maximum (or average) value in each receptive field.
\begin{equation}
    S(i, j)=(I * K)(i, j)=\max_{m, n}\{I(i, j),...,I(i+m, j+n)\}
\end{equation}
Figure \ref{fig:Pooling} shows the pooling operation.

\begin{figure}[H]
\centering
\includegraphics[width=11cm]{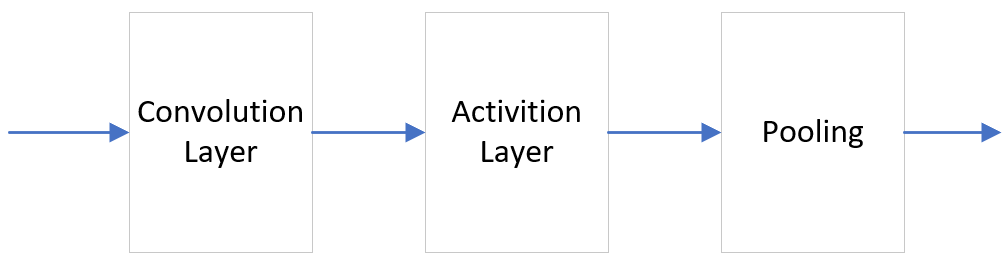}
\caption{CNNs}
\textnormal{CNNs are usually composed with 3 kinds of layer: convolution,
activation and pooling layer.}
\label{fig:CNNs}
\end{figure}

One complete layer in CNNs consists sequentially of convolution, 
activation and pooling layer, as shown in Figure \ref{fig:CNNs}.

Residual CNNs have the same idea as residual DNNs, that map the residual between layers
by adding a skip connection.
One well-known realization is the RESNET presented by \cite{he2016deep}.

\subsection{Sequence Modeling}
\subsubsection{Recurrent Neural Networks}
\begin{figure}[H]
\centering
\includegraphics[width=9cm]{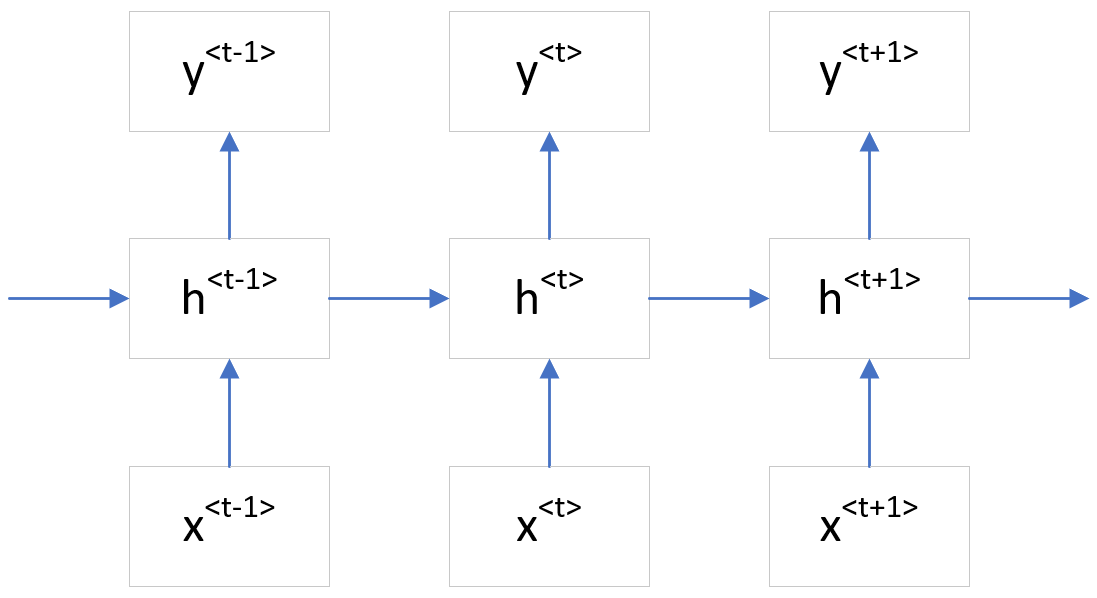}
\caption{RNNs}
\textnormal{RNNs take the input sequence to update the hidden state, which is used to
generate output in each step.}
\label{fig:RNNs}
\end{figure}

RNNs introduced by \cite{rumelhart1985learning} are a family of neural networks
designed for sequential data.
RNNs have a hidden state that keep the state information of each time step,
which are suitable for time series data prediction.
RNNs offer a great way to deal with variable length of input sequence
as shown in Figure \ref{fig:RNNs}.

\begin{figure}[H]
\centering
\includegraphics[height=4.5cm]{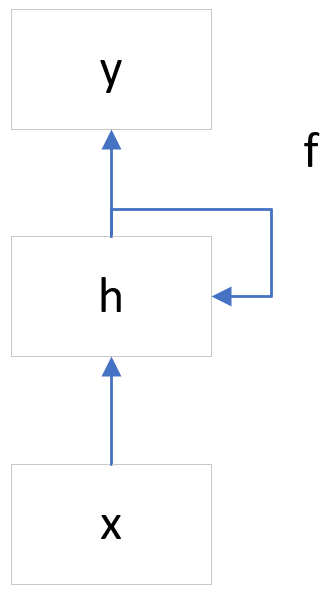}
\caption{Fold Structure of RNNs}
\textnormal{Fold graph of RNNs show the essence of the model that each step in
RNNs share the same structure and parameters.}
\label{fig:RNNs fold}
\end{figure}

RNNs share parameters across all time steps, that the input at any step combines with
the hidden state through the same mechanism (i.e., transition function $f$).
Such structure is illustrated in Figure \ref{fig:RNNs fold}.
\begin{equation}
\boldsymbol{h}^{<t>}=f\left(\boldsymbol{h}^{<t-1>}, \boldsymbol{x}^{<t>} ; 
\boldsymbol{\theta}\right)
\end{equation}
The transition function $f$ consists of two components: affine transformation
and activation function $g_{1}$.
\begin{equation}
h^{<t>}=g_{1}\left(W h^{<t-1>}+U x^{<t>}+b_{1}\right)
\end{equation}
The output $y$ is calculated from hidden state $h$.
\begin{equation}
y^{<t>}=g_{2}\left(V h^{<t>}+b_{2}\right)
\end{equation}
Theoretically, the hidden state will keep most of the predictive information for the output.

\subsubsection{LSTM}
Since RNNs have the same transition function along whole sequence,
the most common problems of RNNs are vanishing and exploding gradient.

Exploding gradient problem can be solved by gradient clipping, which is
proposed by \\
\cite{pascanu2013difficulty}.
It clips the norm $\|g\|$ of the gradient $g$ before parameters are updated.
\begin{equation}
    \begin{aligned}
        \text {if } \| \boldsymbol{g} \| >threshold, \quad
        \boldsymbol{g} \leftarrow \frac{\boldsymbol{g} \star threshold}{\|\boldsymbol{g}\|}
    \end{aligned}
\end{equation}

\begin{figure}[H]
\centering
\includegraphics[width=13cm]{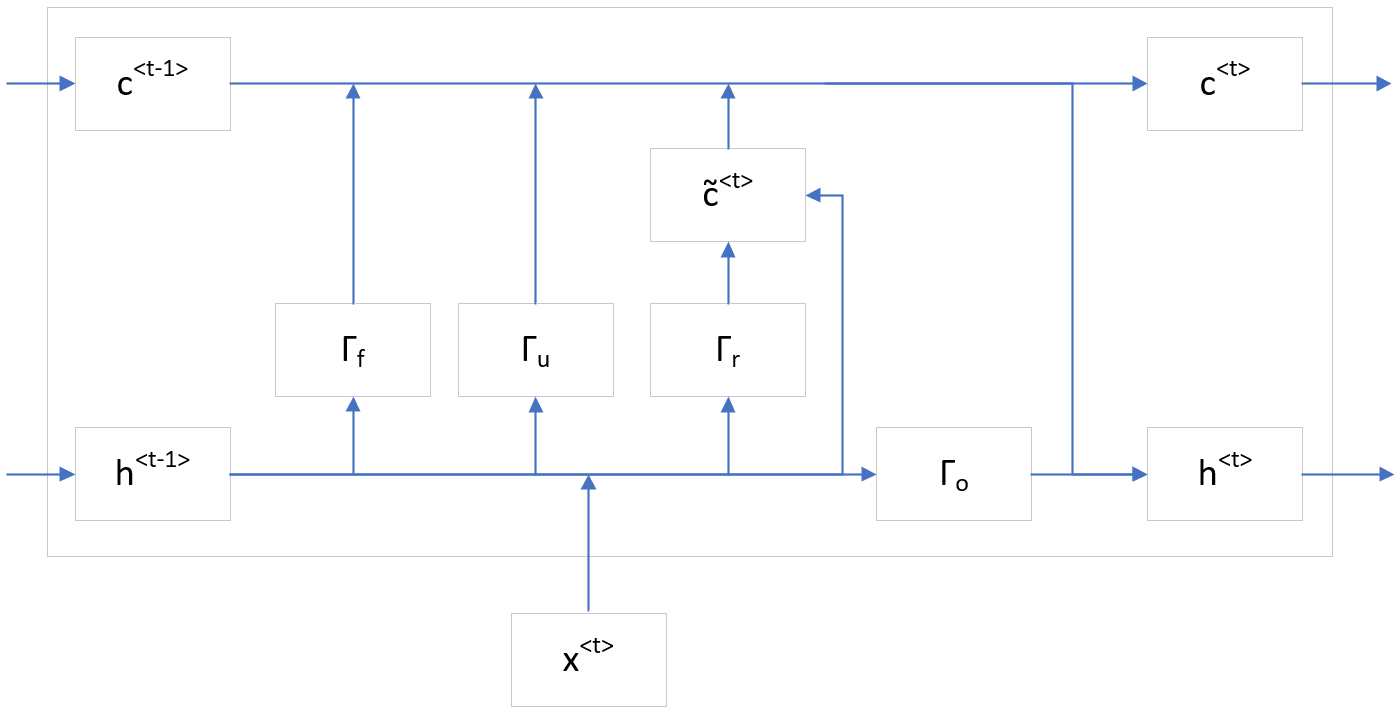}
\caption{LSTM}
\textnormal{LSTM is an enhanced version of RNN with an additional cell unit 
to keep long term memory.}
\label{fig:LSTM}
\end{figure}

Vanishing gradient problem can be solved by adding extra memory in RNNs.
RNNs with such mechanisms are called gated RNNs, including 
LSTM and GRU introduced respectively by
\cite{hochreiter1997long} and \cite{cho2014learning}.
Besides the hidden state, a cell unit is added into the network to keep the memory
of history information along the sequence.
Forget gate and update gate are used to control what information of the inputs
is kept in memory.
The LSTM architecture is shown in Figure \ref{fig:LSTM}.
Therefore, LSTM networks learn long-term dependencies more easily than vanilla RNNs.

The gates are calculated from current input and pervious hidden sate.
\begin{equation}
\Gamma=\sigma\left(W x^{<t>}+U h^{<t-1>}+b\right)
\end{equation}
They are used to control what to forget and what to update in the cell unit.
\begin{equation}
\tilde{c}^{<t>} = 
\tanh \left(W_{c}\left[\Gamma_{r} \star a^{<t-1>}, x^{<t>}\right]+b_{c}\right)
\end{equation}
\begin{equation}
c^{<t>} = 
\Gamma_{u} \star \tilde{c}^{<t>}+\Gamma_{f} \star c^{<t-1>}
\end{equation}
\begin{equation}
h^{<t>} = \Gamma_{o} \star c^{<t>}
\end{equation}

\subsubsection{GRU}
\begin{figure}[H]
\centering
\includegraphics[width=12cm]{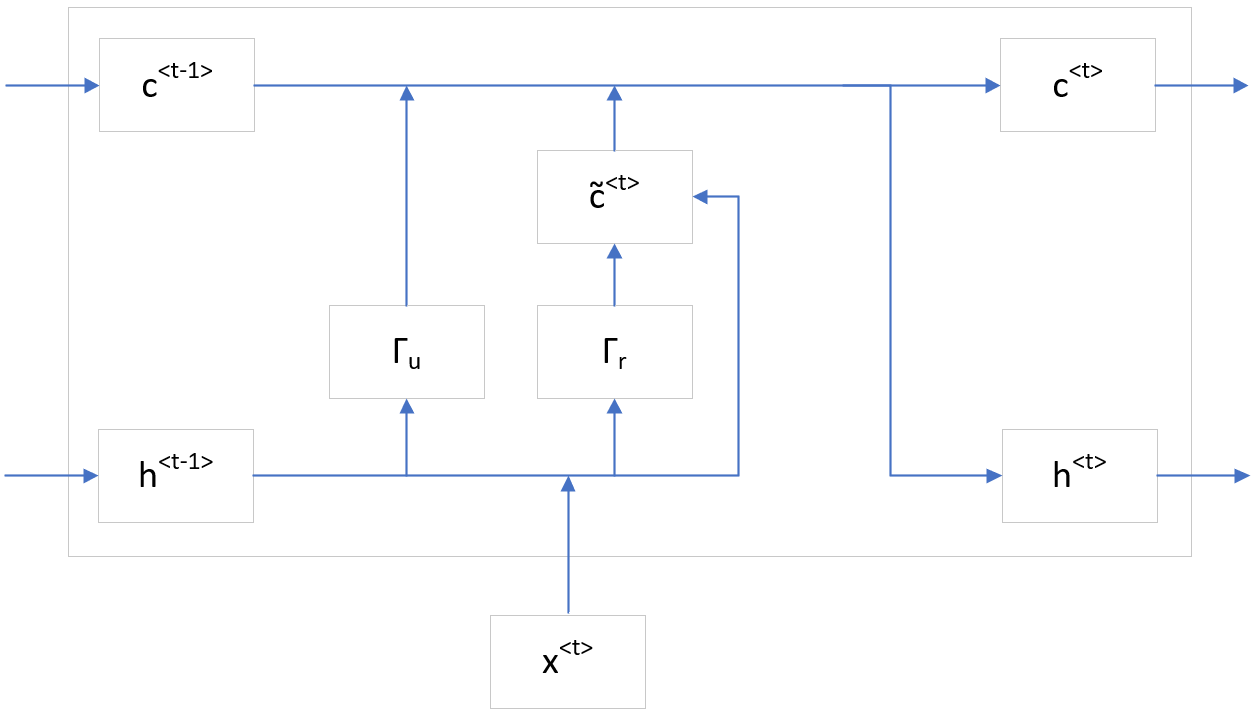}
\caption{GRU}
\textnormal{GRU is a simplified version of LSTM which combine the cell unit and
hidden state together.}
\label{fig:GRU}
\end{figure}

GRU is essentially a simplified version of LSTM, which only use one gate to control both
the forget and update operation. The architecture is shown in Figure \ref{fig:GRU}.
\begin{equation}
\tilde{c}^{<t>} = 
\tanh \left(W_{c}\left[\Gamma_{r} \star a^{<t-1>}, x^{<t>}\right]+b_{c}\right)
\end{equation}
\begin{equation}
c^{<t>} = 
\Gamma_{u} \star \tilde{c}^{<t>}+\left(1-\Gamma_{u}\right) \star c^{<t-1>}
\end{equation}
GRU also merges the information of cell unit and hidden state into one state variable.
\begin{equation}
h^{<t>} = c^{<t>}
\end{equation}

\subsubsection{RNNs with Attention}
Attention mechanism introduced by \cite{bahdanau2014neural} is another method
to solve vanishing gradient. 
The core idea of attention is inspired by attention mechanism in neuroscience.

\begin{figure}[H]
\centering
\includegraphics[width=11cm]{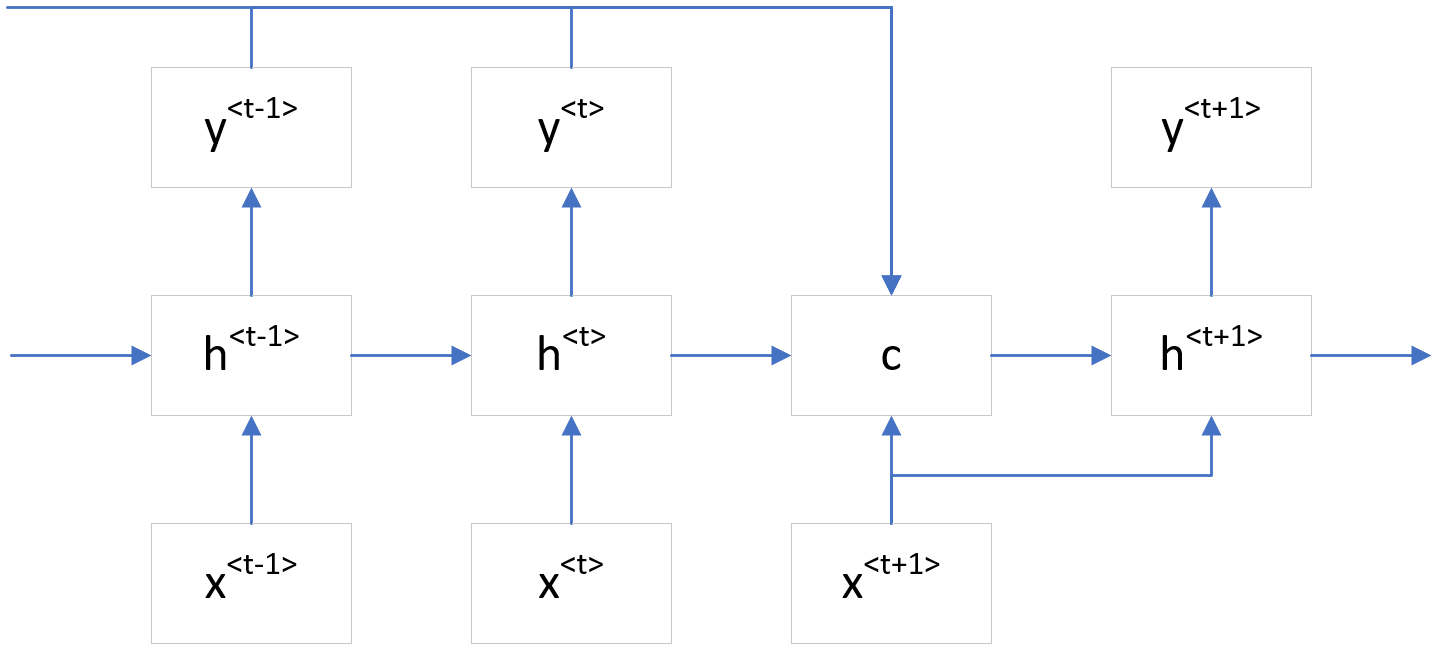}
\caption{RNNs with Attention}
\textnormal{Attention mechanism helps the networks to "pay attention to" history information.}
\label{fig:RNNs with attention}
\end{figure}

From mathematic perspective, attention mechanism is essentially a weighted average of
all pervious information.
A context vector $c$ is constructed by taking a weighted average
of encoder's outputs $y^{<t>}$ as shown in Figure \ref{fig:RNNs with attention}.
In RNNs, the information contained in the outputs are also equivalent to the corresponding
hidden states.
Therefore, the networks will mainly "pay attention to" these memory units 
with most predictive power.

The attention weights $W_{z}$ are first calculated from current input and pervious hidden sate.
\begin{equation}
W_{z} = W_{W}\left[x^{<t+1>}, h^{<t>}\right]+b_{w}
\end{equation}
The context of attention values $c$ is generated from all pervious outputs.
\begin{equation}
z^{<t+1>} = W_{z}Y
\end{equation}
\begin{equation}
c = W_{c}\left[z^{<t+1>}, h^{<t>}\right]
\end{equation}
The decoder will take context $c$ as the hidden layer $h$.
\begin{equation}
h^{<t+1>}=g_{1}\left(W c+U x^{<t+1>}+b_{1}\right)
\end{equation}

\subsubsection{Transformer}
The transformer model introduced by \\
\cite{vaswani2017attention} only emphasizes
on the attention mechanism and gets rid of recurrent structures.
Therefore, Transformer model is entirely based on attention,
and it replaces the recurrent mechanism with multi-headed self-attention.

The model first generates the key $K$, query $Q$ and value $V$ from the inputs.
\begin{equation}
    Q = W_{Q} X
\end{equation}
\begin{equation}
    K = W_{K} X
\end{equation}
\begin{equation}
    V = W_{V} X
\end{equation}
And then calculate the self-attention through $K, Q, V$.
\begin{equation}
Z = \operatorname{Attention}(Q, K, V)=
\operatorname{softmax}\left(\frac{Q K^{T}}{\sqrt{d_{k}}}\right) V
\end{equation}
Each head represents a different kind of attention in every layer.
\begin{equation}
\operatorname{MultiHead}(Q, K, V)=
\operatorname{Concat}\left(\text {head}_{1}, \ldots, \text {head}_{\mathrm{h}}\right) W_{O}
\end{equation}
where
\begin{equation}
\text {head}_{\mathrm{i}}=
\operatorname{Attention}\left(W_{Q, i} Q, W_{K, i} K, W_{V, i} V\right)
\end{equation}

\begin{figure}[H]
\centering
\includegraphics[width=14cm]{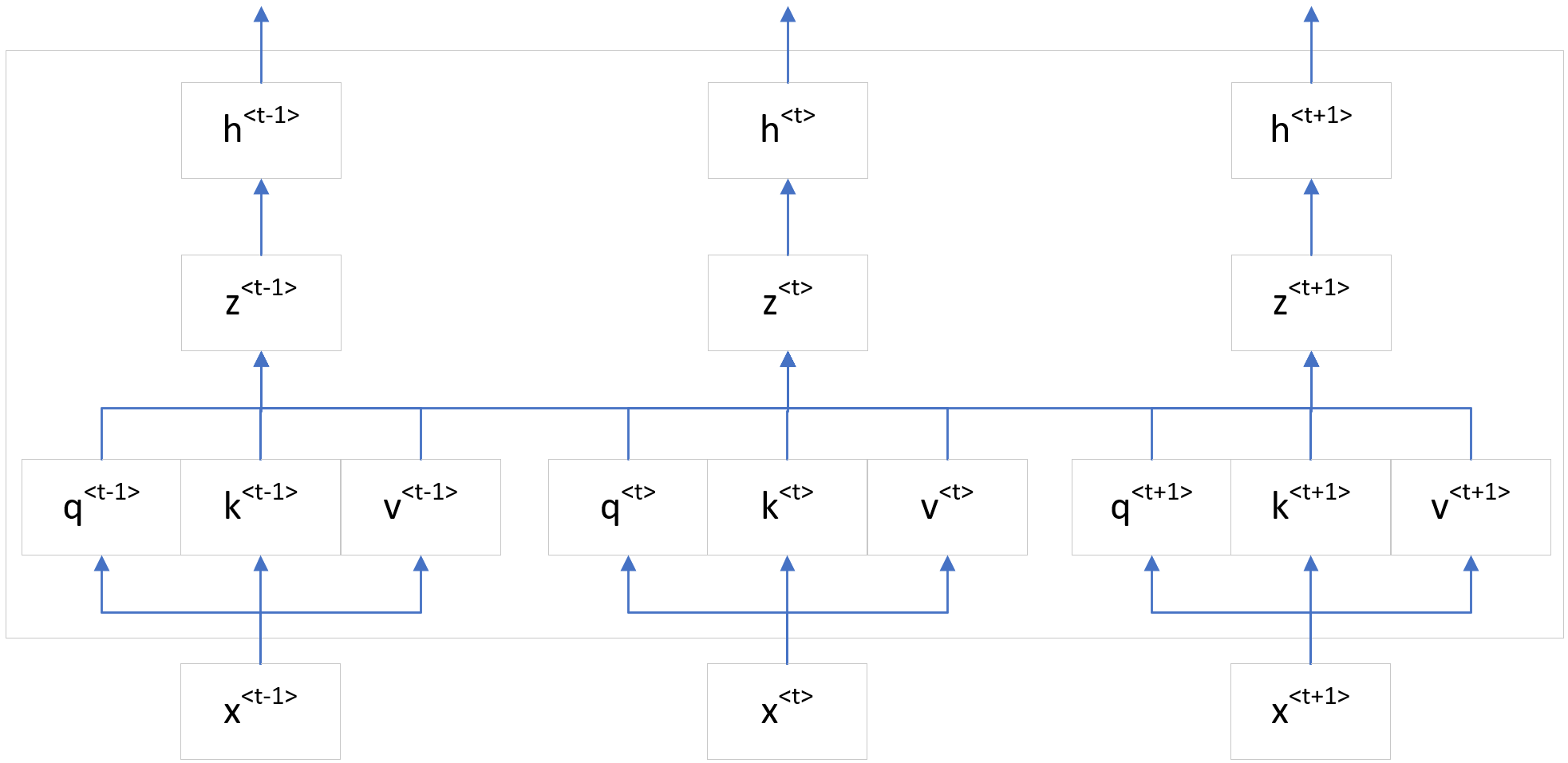}
\caption{Transformer}
\textnormal{Transformer is based on self-attention, it uses multiple layers and multi-head
attentions to explore the deep associations in sequential data.}
\label{fig:Transformer}
\end{figure}

In Transformer, each layer works as transformation of data, which is totally based on
self-attention mechanism applied on all inputs of this layer.
This architecture is shown in Figure \ref{fig:Transformer}.

\subsection{Performance Evaluation}
I use out-of-sample $R^{2}$ to evaluate the performance of all my models.
I use the same $R^{2}$ metric as \\
\cite{gu2020empirical} proposed and justified,
whose denominator is the sum of squared returns without demeaning.
\begin{equation}
R_{\mathrm{oos}}^{2}=1-\frac{\sum_{(i, t) \in \mathcal{T}}\left(r_{i, t+1}-
\widehat{r}_{i, t+1}\right)^{2}}{\sum_{(i, t) \in \mathcal{T}} r_{i, t+1}^{2}}
\end{equation}
$\mathcal{T}$ represents for test data set, which is only used for the final
performance evaluation. 
Therefore, I focus on the out-of-sample performance during my whole analysis on
all deep learning methods.

\section{Monte Carlo Simulations}
I use a Monte Carlo simulation algorithm with similar parameters calibrated by \cite{gu2020empirical},
For each Monte Carlo sample, I divide the whole time series into 3 consecutive subsamples of
80\% for training, 10\% for validation and 10\% for testing, respectively.

I estimate both linear and nonlinear model in the training sample, using 
Random Forest (RF), Gradient Boosted Regression Trees (GBRT), LightGBM (LGBM),
Multilayer Perceptron (MLP), Multilayer Perceptron (MLP) + Residual
Convolutional Neural Networks (CNNs), Recurrent Neural Network (RNN),
Recurrent Neural Network (RNN) + Attention, Gated Recurrent Unit (GRU)
Long Short-Term Memory (LSTM) and Transformer.
I choose tuning parameters for each method in the validation sample, and calculate
the prediction errors in the testing sample. For benchmark, I also compared OLS and oracle model.

\begin{table}[H]
\caption{Comparison of Predictive $R^{2}$s for Machine Learning Algorithms in Simulations}
\centering
\begin{tabular}{lccccccccccc}
\hline
Model        &&\multicolumn{3}{c}{Linear}&&   &&\multicolumn{3}{c}{Nonlinear}\\
\hline
$R^{2}(\%)$  && IS      && OOS       &&       && IS     && OOS     \\
\hline        
OLS          && 4.23    && 3.38      &&       && 3.32   && 2.32    \\
RFR          && 7.37    && 1.45      &&       && 9.06   && 3.91    \\
GBRT         && 6.69    && 4.19      &&       && 8.52   && 5.64    \\
LGBMR        && 6.02    && 4.34      &&       && 8.53   && 5.68    \\
MLP          && 3.61    && 2.69      &&       && 4.92   && 3.27    \\
MLP+R        && 3.88    && 2.91      &&       && 5.51   && 4.57    \\
CNN          && 4.12    && 3.29      &&       && 11.84  && 2.10    \\
RNN          && 3.66    && 2.63      &&       && 3.78   && 2.91    \\
RNN+A        && 3.78    && 2.67      &&       && 5.73   && 4.32    \\
GRU          && 4.34    && 2.71      &&       && 6.04   && 4.13    \\
LSTM         && 3.68    && 2.73      &&       && 4.35   && 3.58    \\
Transformer  && 3.44    && 2.67      &&       && 5.86   && 4.14    \\
Oracle       && 5.22    && 7.23      &&       && 7.88   && 7.71    \\
\hline
\end{tabular}\label{tbl:Simulation Table}
    \includegraphics[width=18cm]{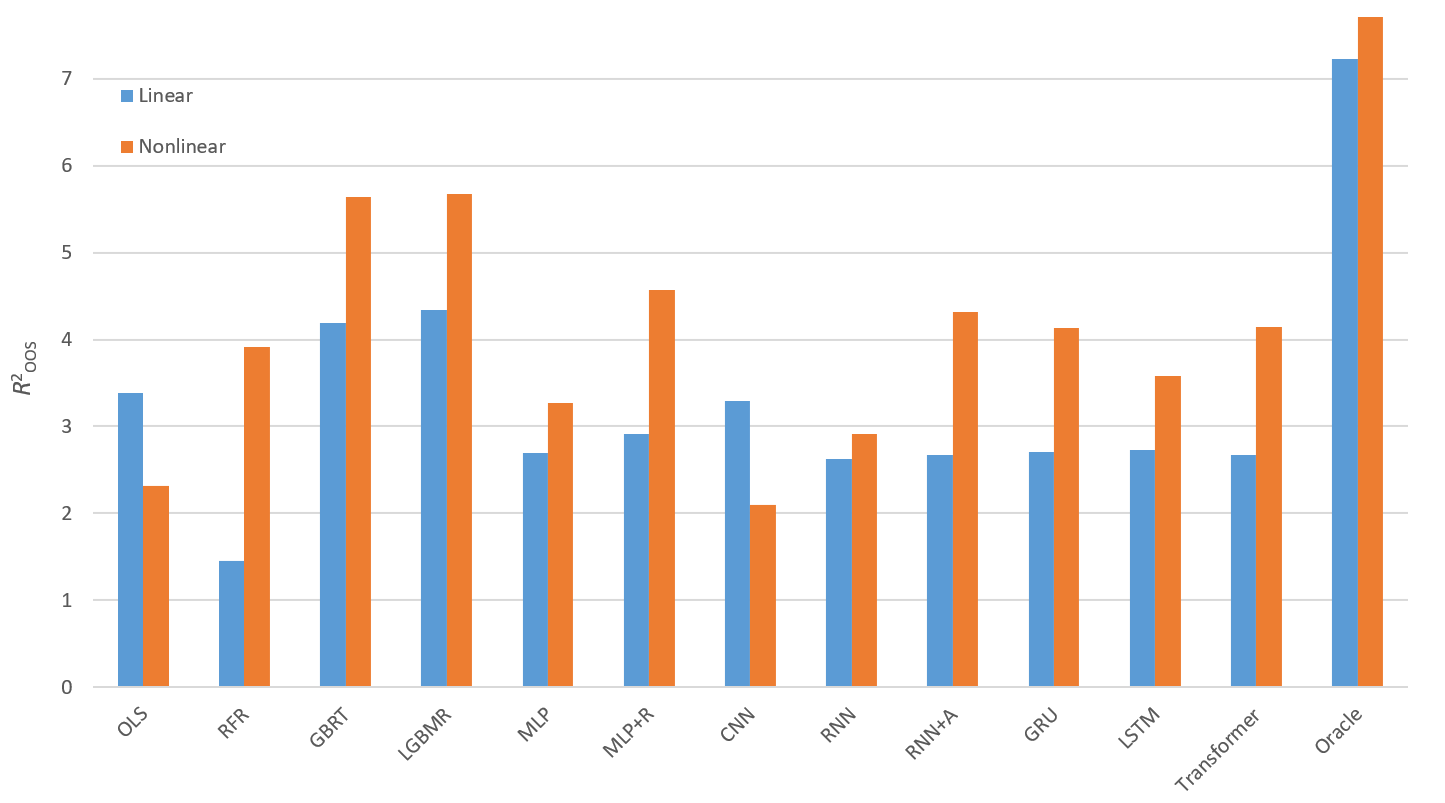}
\end{table}

I report the average $R^{2}$s both in-sample (IS) and out-of-sample (OOS) for each model and
each method over 10 Monte Carlo repetitions in Table \ref{tbl:Simulation Table}.

For linear model, OLS and advanced tree methods deliver the best
out-of-sample $R^{2}$. 
By contrast, for nonlinear model, these methods clearly
dominate OLS, because the latter cannot capture the nonlinearity in the model.
NNs is better, but is dominated by RF, and GBRT. I think that's because traditional
ML methods are good at simple nonlinear tasks, while NNs are specialized in more complicated cases.
Especially, NNs are best performed methods in nonstructural and complicated scenarios.
OLS is the worst in all settings, not surprisingly.

\section{An Empirical Study of US Equities}
\subsection{Data and Over-arching Model}
I take all CRSP firms listed on the NYSE, AMEX, or NASDAQ as my samples.
I obtain their monthly equity returns from July 1976 to December 2016,
and the total time span is 40 years. 
I construct my empirical samples with about 28,000 stocks totally
and 6,000 per month on average.

I take the one-month Treasury bill rate from Kenneth R. French' data library to
proxy for the risk-free rate. 
Then, I calculate the excess return by using stock return minus risk-free rate.

I construct 81 stock-level characteristics based on the cross section of stock returns
literature, that mainly follows the work of \cite{green2017characteristics}.

I construct 5 factors proposed by \cite{fama2015five} as proxies for systematic risk.

I also construct 6 macroeconomic proxies according to the work of \cite{welch2008comprehensive},
including dividend-price ratio (dp), earnings-price ratio (ep), book-to-market
ratio (bm), net equity expansion (ntis), Treasury-bill rate (tbl) and stock variance (svar).

Just like the work of \cite{gu2020empirical} for machine learning, all
my deep learning methods are designed to approximate
the over-arching empirical model $E_{t}(r_{i, t+1})=f^{\star}(x_{i, t})$,
which is defined in equation (\ref{eq:over-arching}), and
baseline set of stock-level covariates $z_{i,t}$ is defined as
\begin{equation}
    z_{i, t}=x_{t} \otimes c_{i, t}
\end{equation}
The theory foundation for this model, as mentioned by \cite{gu2020empirical},
is the standard beta-pricing representation of the asset pricing conditional Euler equation.
Stock-level characteristics $c_{i,t}$ are used in analogy with risk exposure function
$\beta_{i,t}$, systemic risk and macroeconomic proxies are used in analogy with
risk premium $\lambda_{t}$.
Let $\beta_{i, t}=\theta_{1} c_{i, t}, \lambda_{t}=\theta_{2} x_{t}$,
\begin{equation}
\mathrm{E}_{t}\left(r_{i, t+1}\right)=\beta_{i, t}^{\prime} \lambda_{t}
\end{equation}
\begin{equation}
    \begin{aligned}
        g^{\star}\left(z_{i, t}\right) & = \mathrm{E}_{t}\left(r_{i, t+1}\right) \\
        & = \beta_{i, t}^{\prime} \lambda_{t} \\
        & = c_{i, t}^{\prime} \theta_{1}^{\prime} \theta_{2} x_{t} \\
        & = \left(x_{t} \otimes c_{i, t}\right)^{\prime} 
        \operatorname{vec}\left(\theta_{1}^{\prime} \theta_{2}\right) \\
        & =: z_{i, t}^{\prime} \theta,
    \end{aligned}
\end{equation}
where
\begin{equation}
    \theta=\operatorname{vec}\left(\theta_{1}^{\prime} \theta_{2}\right)
\end{equation}

The Deep learning model is more general, because the $g^{\star}(z_{i, t})$ 
can be any function and the input $z_{i, t}$ can be any combination of
$x_{t}$ and $c_{i, t}$.
Therefore, the factor risk premium can have nonlinear relationship with macroeconomic
and systematic risk, $\lambda_{t}=f_{1}(x_{t})$.
Also the risk exposure can have nonlinear relationship with firm characteristics,
$\beta_{i, t}= f_{2}(c_{i, t})$.

I design 2 kinds of time span samples, and separate the data set as follows. 

First, long time span contains totally 40 years.
80\% for train set (1976 - 2007);
10\% for validation set (2008 - 2012);
10\% for test set (2013 - 2016).

Second, short time span contains totally 5 years.
80\% for train set (2011.12 - 2016.06);
10\% for validation set (2016.06 - 2016.12);
10\% for test set (2016.07 - 2016.12).

\subsection{The Cross Section of Individual Stocks}
Since the work of \cite{gu2020empirical} and \cite{LEIPPOLD2021} have both conduct 
a comparative analysis on various machine learning methods and identified basic DNNs
as the best preforming model, I focus on SOTA deep learning models
and start with basic DNNs.

I compare 8 deep learning models in total, including OLS, DNNs, residual DNNs, CNNs,
residual CNNs, RNNs, RNNs with attention, GRU, LSTM and Transformer.
Table \ref{tbl:Summary Table} presents the comparison of different deep learning models
in terms of their out-of-sample $R^{2}$. 
I use both long time span (40 years) and short time span (5 years) samples as the data set.

\begin{table}[H]
    \centering
        \caption{Monthly Out-of-sample Stock-level Prediction Performance (Percentage $R^{2}_{OOS}$)}
        \begin{tabular}{ccccccccccccc}
            \hline
                  & OLS & RFR & GBRT & LGBMR & MLP & MLP & CNN & RNN & RNN & GRU & LSTM & Transformer \\
                  &  &  &  &  &  & +R &  &  & +A &  &  &  \\
            \hline
            20 Years  & -0.50 & 1.05 & 0.53 & 0.63 & 1.14 & 1.25 & 1.00 & 1.06 & 1.38 & 1.44 & 1.44 & 1.34 \\
            \hline
        \end{tabular}\label{tbl:Summary Table}
    \includegraphics[width=18cm]{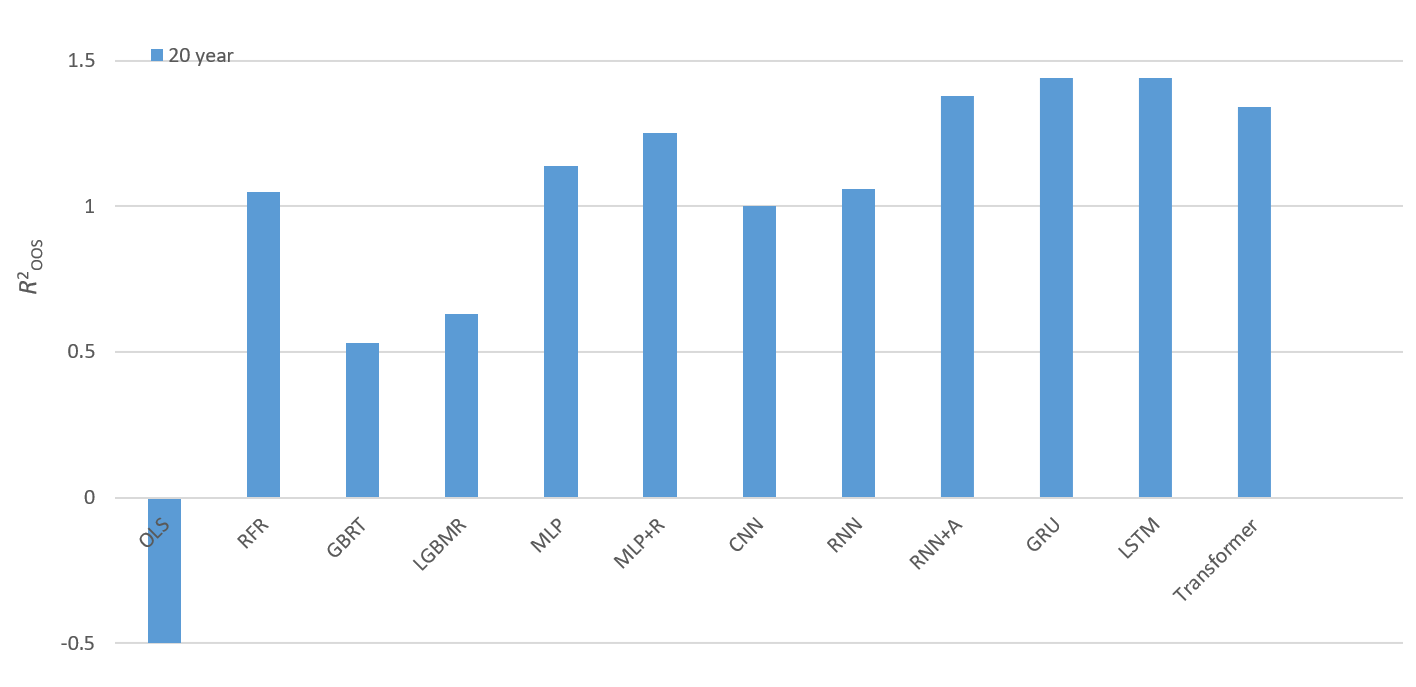}
\end{table}

The main results of my comparative analysis are as follows.

First, Compare with traditional linear regression method, all SOTA deep learning methods have
far better performance in stock return prediction.

Second, RNNs have better performance.
This demonstrates that past information have predictive power on present return prediction,
and the underlying correlation exists between today's return and history data.
The high performance also justifies that RNNs are good at merging past and
present information together to make predictions.
However, vanilla RNNs in long time span data set are not trainable due to
the vanishing gradient problem.

Third, the memory mechanisms of attention and cell unit in RNNs both work well. 
It highlight the value of long term memory in stock return prediction.

Forth, GRU as a simplify version of LSTM, make the model much simpler and without loss of
performance at the same time.
According to Occam's Razor principle, GRU could be a excellent substitution of LSTM 
in empirical asset pricing.

Fifth, CNNs have relative worse performance than other NNs.
It's due to the design of CNNs is focusing on the Computer Vision (CV) tasks instead of 
stock return prediction.
What's more, Transformer works always better than RNNs in most NLP tasks,
while it is not the case in stock return prediction.
These findings demonstrate that a good design is mainly based on the domain knowledge,
and highlight the importance of prior knowledge and domain theory guided model design.

Last, deeper NNs with skip connections have only a little improvement.
This suggests that the NNs needn't much the depth, 
and instead middle or shallow networks can also work well.
Therefore, I should not overestimate the depth and complexity of
the fundamental associations between market data and stock returns, and could be 
more optimistic about the future research of asset pricing theory.
Learning from data and then to figure out the underlying economic mechanisms
through explainable AI methods could be another promising research method,
which can promote the study of asset pricing.

\begin{table}[H]
\caption{Diebold-Mariano Tests of Out-of-Sample Prediction}
\centering
\begin{tabular}{l|ccccccccccc}
\hline
& RFR & GBRT & LGBMR & MLP & MLP & CNN & RNN & RNN & GRU & LSTM & Transformer \\
&  &  &  &  & +R &  &  & +A &  &  &  \\
\hline        
OLS          &\bftab 20.64&\bftab 13.83&\bftab 17.4&\bftab 21.35&\bftab 23.83&\bftab 28.32&\bftab 22.08&\bftab 26.74&\bftab 26.86&\bftab 27.22&\bftab 25.9\\
RFR          &&\bftab -23.35&\bftab -15.89&\bftab -3.19&     1.49&\bftab -7.33&\bftab -5.84&\bftab 7.21&\bftab 8.54&\bftab 7.09&\bftab 4.3\\
GBRT         &&&\bftab 15.08&\bftab 12.61&\bftab 10.68&      1.75&\bftab 7.21&\bftab 22.08&\bftab 24.73&\bftab 19.24&\bftab 20.65\\
LGBMR        &&&&\bftab 8.46&\bftab 7.14&       -1.92&\bftab 2.46&\bftab 16.93&\bftab 19.82&\bftab 16.57&\bftab 15.43\\
MLP          &&&&&\bftab 3.38&\bftab -6.54&\bftab -3.54&\bftab 10.37&\bftab 12.49&\bftab 13.13&\bftab 8.03\\
MLP+R        &&&&&&\bftab -11.29&\bftab -7.04&\bftab 4.24&\bftab 4.26&\bftab 3.71&       1.29\\
CNN          &&&&&&&\bftab 4.66&\bftab 15.44&\bftab 15.33&\bftab 15.25&\bftab 13.01\\
RNN          &&&&&&&&\bftab 15.89&\bftab 17.36&\bftab 15.55&\bftab 13.16\\
RNN+A        &&&&&&&&&       0.69&       0.35&\bftab -6.7\\
GRU          &&&&&&&&&&       -0.12&\bftab -8.79\\
LSTM         &&&&&&&&&&&\bftab -5.6\\
\hline
\end{tabular}\label{tbl:DM Table}
\begin{flushleft}
\end{flushleft}
\end{table}

Table \ref{tbl:DM Table} assesses the statistical significance of thirteen models,
what reports the pairwise Diebold-Mariano test statistics of thirteen models.
Positive statistic indicate the column model outperforms the row model and bold numbers
indicate significance at the $5\%$ level.
The results consist with pervious discussion, 
and RNN with memory units and attention mechanisms have best performance.


\subsection{Portfolio Forecasts}
I conduct portfolios experiments according to machine learning forecasts by
calculating one-month-ahead out-of-sample stock return predictions.
I first sort stocks into deciles based on prediction of all 13 models,
and then reconstitute portfolios each month with a zero-net-investment portfolio
long the highest decile and short the lowest one.

Table \ref{tbl:Port Table} reports the performance of portfolios over the 5-year testing period.
Column "Pred", "Avg", "Std", and "SR" represent the predicted monthly returns,
and realized average monthly returns, standard deviations and Sharpe ratios.
All portfolios are equally weighted.
The results are consistent with pervious deep learning forecast performance, 
except that OLS has a relative good performance comparing to its OOS $R^{2}$.
Realized returns almost consist with deep learning forecasts,
and RNN models with memory mechanisms still dominate linear models and
traditional ML (tree-based) approaches.

\begin{table}[H]
\renewcommand\arraystretch{0.98}
    \centering
        \caption{Performance of Machine Learning Portfolios}
        \begin{tabular}{ccccccccccccccc}
\hline
&\multicolumn{4}{c}{OLS}& &\multicolumn{4}{c}{RFR}&& \multicolumn{4}{c}{GBRT} \\
\hline
        & Pred & Avg & Std & SR && Pred & Avg & Std & SR && Pred & Avg & Std & SR \\
\hline
Low(L)  &-3.78&0.3&7.31&0.14        &&-3.38&1.13&7.58&0.52    &&-0.08&1.11&5.89&0.65 \\
2       &-2.23&0.41&5.63&0.25       &&-2.38&1.33&6.75&0.68    &&0.05&1.08&5.72&0.65 \\
3       &-1.7&0.69&5.26&0.45        &&-1.91&1.5&6.05&0.86     &&0.15&1.36&5.94&0.79 \\
4       &-1.32&0.74&4.88&0.52       &&-1.47&1.46&5.82&0.87    &&0.22&1.13&5.93&0.66 \\
5       &-1.02&1.23&4.66&0.91       &&-1.07&0.98&5.19&0.65    &&0.22&0.84&4.82&0.6 \\
6       &-0.73&1.2&4.74&0.88        &&-0.72&0.9&4.79&0.65     &&0.22&1.17&4.84&0.84 \\
7       &-0.44&1.48&4.85&1.05       &&-0.39&1.33&4.98&0.93    &&0.23&0.81&4.55&0.62 \\
8       &-0.1&1.58&4.97&1.1         &&-0.04&1.38&4.86&0.98    &&0.29&2.05&5.95&1.19 \\
9       &0.37&2.4&5.78&1.44         &&0.44&1.48&5.69&0.9      &&0.4&1.7&5.78&1.02 \\
High(H) &1.55&4.54&7.64&2.06        &&2.3&3.06&7.48&1.42      &&0.83&3.31&7.5&1.53 \\
H-L     &5.32&4.24&4.43&3.31        &&5.68&1.93&7.25&0.92     &&0.91&2.21&4.14&1.84 \\
\hline
&\multicolumn{4}{c}{LGBMR}& &\multicolumn{4}{c}{MLP}&& \multicolumn{4}{c}{MLP+R} \\
\hline
        & Pred & Avg & Std & SR && Pred & Avg & Std & SR && Pred & Avg & Std & SR \\
\hline
Low(L)  &0.39&1.26&5.9&0.74     &&0.33&0.2&7.16&0.1          &&-0.62&-0.04&7.94&-0.02 \\
2       &0.56&1.08&5.8&0.64     &&0.79&0.34&5.36&0.22        &&0.02&0.15&6.39&0.08 \\
3       &0.62&1.42&6.02&0.82    &&1.04&0.76&6&0.44           &&0.3&0.53&5.57&0.33 \\
4       &0.67&1.31&6.42&0.71    &&1.22&0.62&5.25&0.41        &&0.51&0.79&5.12&0.53 \\
5       &0.67&1.25&5.58&0.78    &&1.39&0.9&4.92&0.63         &&0.71&0.9&4.86&0.64 \\
6       &0.67&1.1&4.76&0.8      &&1.55&1.33&4.65&0.99        &&0.92&1.19&4.85&0.85 \\
7       &0.68&1.13&4.84&0.81    &&1.7&1.53&5.14&1.03         &&1.15&1.69&4.79&1.22 \\
8       &0.7&1.98&6.32&1.08     &&1.86&1.77&5.09&1.2         &&1.43&1.78&4.89&1.26 \\
9       &0.75&1.78&5.72&1.08    &&2.03&2.9&5.83&1.72         &&1.85&2.49&5.24&1.65 \\
High(H) &1.01&2.25&5.66&1.38    &&2.4&4.22&7.34&1.99         &&2.96&5.09&7.9&2.23 \\
H-L     &0.62&0.99&3.54&0.97    &&2.07&4.02&4.63&3.01        &&3.57&5.12&5.38&3.3 \\
\hline
&\multicolumn{4}{c}{CNN}& &\multicolumn{4}{c}{RNN}&& \multicolumn{4}{c}{RNN+A} \\
\hline
        & Pred & Avg & Std & SR && Pred & Avg & Std & SR && Pred & Avg & Std & SR \\
\hline
Low(L)  &-1.47&-0.45&7.16&-0.22  &&-0.61&-0.07&8.24&-0.03   &&-0.78&0.02&7.99&0.01 \\
2       &-0.47&0.18&5.84&0.11    &&0.16&-0.03&6.17&-0.02    &&-0.06&0.37&6.23&0.2 \\
3       &-0.11&0.43&5.5&0.27     &&0.5&0.31&5.66&0.19       &&0.31&0.48&5.59&0.3 \\
4       &0.15&0.79&4.95&0.55     &&0.74&0.4&5&0.28          &&0.58&0.42&4.89&0.3 \\
5       &0.38&1.11&4.99&0.77     &&0.94&0.81&4.62&0.61      &&0.8&0.75&4.58&0.57 \\
6       &0.59&1.26&4.65&0.94     &&1.12&1.4&4.5&1.08        &&1&1.22&4.57&0.92 \\
7       &0.82&1.61&4.79&1.17     &&1.31&1.76&4.37&1.39      &&1.22&1.69&4.38&1.34 \\
8       &1.1&2.1&5.11&1.42       &&1.54&2.09&5.05&1.43      &&1.48&2.05&4.94&1.44 \\
9       &1.47&2.58&5.6&1.59      &&1.86&2.88&5.67&1.76      &&1.86&2.72&6.09&1.55 \\
High(H) &2.48&4.95&7.79&2.2      &&2.6&5&8.09&2.14          &&2.76&4.83&8.12&2.06 \\
H-L     &3.95&5.41&5.09&3.68     &&3.21&5.07&4.34&4.05      &&3.54&4.8&5.06&3.29 \\
\hline
&\multicolumn{4}{c}{GRU}& &\multicolumn{4}{c}{LSTM}&& \multicolumn{4}{c}{Transformer} \\
\hline
    & Pred & Avg & Std & SR && Pred & Avg & Std & SR && Pred & Avg & Std & SR \\
\hline
Low(L)  &-1.04&-0.23&7.87&-0.1 &&-0.63&-0.26&7.88&-0.11   &&-0.55&-0.07&7.88&-0.03 \\
2       &-0.24&0.2&6.76&0.1    &&0.11&0.06&6.08&0.03      &&0.12&0.05&5.89&0.03 \\
3       &0.23&0.44&5.51&0.28   &&0.49&0.41&5.41&0.27      &&0.45&0.35&5.63&0.22 \\
4       &0.56&0.57&5.07&0.39   &&0.75&0.46&5&0.32         &&0.7&0.68&4.99&0.47 \\
5       &0.81&0.93&4.83&0.67   &&0.95&0.72&4.71&0.53      &&0.9&1.2&4.75&0.88 \\
6       &1.03&1.15&4.45&0.89   &&1.14&1.41&4.64&1.06      &&1.09&1.58&4.65&1.18 \\
7       &1.25&1.69&4.5&1.3     &&1.33&1.69&4.61&1.27      &&1.28&1.84&4.75&1.34 \\
8       &1.54&2.05&4.89&1.45   &&1.58&2.21&5.09&1.51      &&1.49&1.97&4.9&1.39 \\
9       &1.95&2.73&5.5&1.72    &&1.95&2.8&5.98&1.62       &&1.78&2.71&5.67&1.66 \\
High(H) &2.86&5.03&8.26&2.11   &&2.88&5.05&8.49&2.06      &&2.32&4.25&7.83&1.88 \\
H-L     &3.9&5.26&4.63&3.94    &&3.51&5.31&5.22&3.53      &&2.87&4.32&4.24&3.53 \\
\hline
\end{tabular}\label{tbl:Port Table}
\begin{flushleft}
\end{flushleft}
\end{table}

\begin{figure}[H]
\centering
\includegraphics[width=18cm]{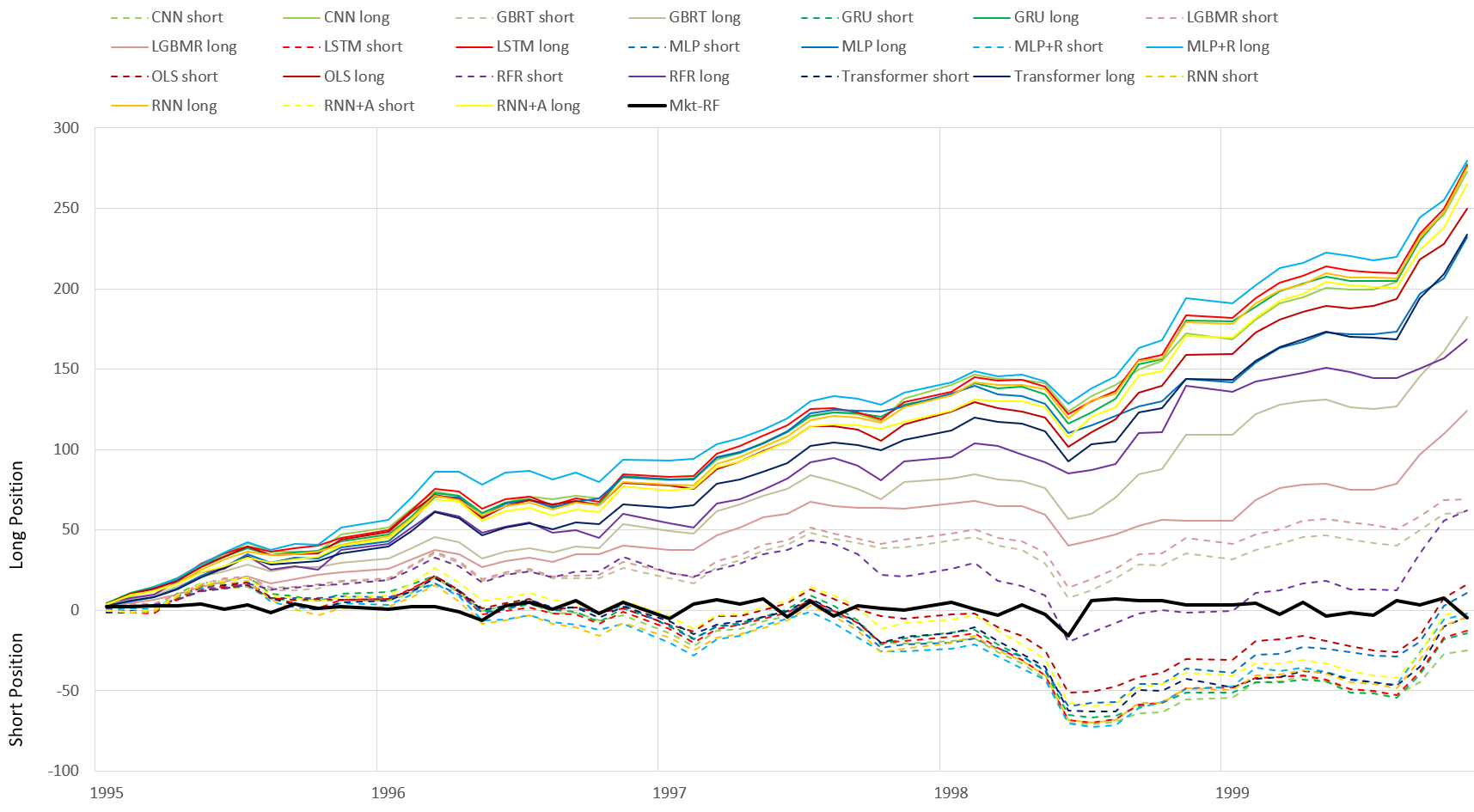}
\caption{Cumulative Return of Machine Learning Portfolios}
\begin{flushleft}
\end{flushleft}
\label{fig:ML port}
\end{figure}

Figure \ref{fig:ML port} reports
cumulative log returns for long and short sides of portfolios sorted on models' forecasts.
Bold black baseline is cumulative market excess return.
The solid and dash lines represent long (top decile) and short (bottom decile) positions.
Finally, RNN with memory mechanisms and Residual MLP dominate all other models in both directions.

\section{Conclusion}
I conduct a comprehensive comparative analysis of various deep learning methods on 
empirical asset pricing, and identify that RNNs with memory mechanism and Transformer
have the best performance in terms of predictivity. 
Furthermore, I demonstrate large economic gains to investors using deep learning forecasts.

My comparisons of various NNs' performance on stock return prediction
demonstrate the importance of domain knowledge and financial theory when
designing deep learning models. 
This finding inspires a promising research direction on empirical asset pricing
via deep learning in future work .
My findings show prediction in asset pricing brings new challenges to deep learning,
that the samples violate IID assumption accepted by most deep learning applications.
The distribution shift problem caused by time varying distribution and
the low signal-to-noise ratio problem of financial data are the two urgent problems
need solving in stock return prediction models.
The finding that well performed networks are not always deep
gives a promising direction for future study of asset pricing theory.
Learning from data and figuring out the insights of asset pricing theory behind deep
learning models through explainable AI theory is a promising research method.

The success of deep learning methods for stock return prediction brings great prospects for
innovative economic models, and highlights the value of deep learning in empirical
understanding of asset prices. 
What's more, it also suggests that all the follow-up achievements in the booming
deep learning studies (e.g., explainable AI theory, innovative architectures, mechanisms
and models, etc.) can constantly promote the study of asset pricing.
The better measurement with deep learning methods can improve the risk premia prediction,
and promote and simplify the study of underlying economic mechanisms behind asset pricing.

Overall, my findings not only justify the role of deep learning methods in the booming
financial innovation, but also highlight their prospects and advantages over
traditional machine learning methods.

\bibliographystyle{rfs}
\bibliography{apdl}

\end{document}